# MSPINN: Multiple scale method integrated physics-informed neural networks for reconstructing transient natural convection

Nagahiro Ohashi[1], Nam Phuong Nguyen[1], Leslie K. Hwang[2] and Beomjin Kwon[1]

[1]School for Engineering of Matter, Transport and Energy, Arizona State University, Tempe, AZ 85287, USA

[2]School of Electrical, Computer and Energy Engineering, Electrical Engineering, Arizona State University, Tempe, AZ 85287, USA

*Corresponding author: kwon@asu.edu

**Abstract**

This study employs physics-informed neural networks (PINNs) to reconstruct multiple flow fields in a transient natural convection system solely based on instantaneous temperature data at an arbitrary moment. Transient convection problems present reconstruction challenges due to the temporal variability of fields across different flow phases. In general, large reconstruction errors are observed during the incipient phase, while the quasi-steady phase exhibits relatively smaller errors, reduced by a factor of 2 to 4. We hypothesize that reconstruction errors vary across different flow phases due to the changing solution space of a PINN, inferred from the temporal gradients of the fields. Furthermore, we find that reconstruction errors tend to accumulate in regions where the spatial gradients are smaller than the order of $10^{-6}$, likely due to the vanishing gradient phenomenon. In convection phenomena, field variations often manifest across multiple scales in space. However, PINN-based reconstruction tends to preserve larger-scale variations, while smaller-scale variations become less pronounced due to the vanishing gradient problem. To mitigate the errors associated with vanishing gradients, we introduce a multi-scale approach that determines scaling constants for the PINN inputs and reformulates inputs across multiple scales. This approach improves the maximum and mean errors by 72.2% and 6.4%, respectively. Our research provides insights into the behavior of PINNs when applied to transient convection problems with large solution space and field variations across multiple scales.

# 1. Introduction

Limitation in data availability promoted the development of physics-informed neural networks (PINNs), wherein the underlying physics of the problem are integrated into the neural network architecture. PINNs address the data limitation issue by inherently learning to approximate the solutions to partial differential equations (PDEs) that characterize the physics of the system. Unlike traditional machine learning methods that rely on large datasets, PINNs leverage the fundamental governing equations of the system to guide the learning process. Pioneering work by Raissi et al. introduced PINNs to address forward and inverse nonlinear problems, using Burger and Schrodinger equations [1]. This deep learning framework has been successful in applications across various engineering disciplines, as reviewed by Cuomo et al. [2]. In the study of forward thermofluid problems, Cai et al. provided concise overviews of common applications demonstrating the predictive capabilities of PINNs across diverse convection phenomena, including forced and mixed convection with partially known boundary conditions [3]. Additionally, Lucor et al. used PINN to predict temperature and velocity fields of turbulent Rayleigh-Benard convection within both rough and smooth rectangular cavities at a Rayleigh number (Ra) of $2\times10^7$ [4]. However, their research also highlighted two key limitations of the standard PINN configuration: 1) Reduced accuracy in regions near spatial and temporal boundaries within the domain, and 2) Inaccuracy in handling complex dynamics with sharp field gradients. To address these challenges, they implemented two strategies: 1) Expansion of the original domain through temporal and spatial padding, and 2) Relaxation of the most sensitive constraint in Navier-Stokes equation set, specifically the pressure field constraint imposed by incompressibility constraint.

Another important application of PINNs is in solving inverse problems, such as reconstructing physical fields. Flow field reconstruction is an important yet challenging task in thermofluid research, that is the process of estimating velocity, temperature or other flow variables throughout a fluid domain based on partial or indirect measurements. In many real-world applications, obtaining complete flow conditions, including velocity, temperature, and pressure distributions at all flow inlets, outlets, and other boundaries, is often impractical or impossible. For instance, laser diagnostics like particle image velocimetry, which measure continuous fluid flow velocity distributions, require light to pass through the fluid system. However, most real-world flow systems are constructed from metals with opaque surfaces, making it challenging to measure complete velocity distributions. Additionally, while point measurement sensors can be inserted into opaque flow systems, they tend to disturb the original fluid flows. Traditional methods that compute the flow fields from boundary conditions, such as computational fluid dynamics (CFD), typically demand a fully defined system. To address these challenges in flow field reconstruction, PINNs have recently attracted significant interest for their ability to provide comprehensive field information from limited measurement data.

There have been initial investigations into the application of PINNs for reconstructing thermofluidic fields. Wang et al. reconstructed a two-dimensional (2D) natural convection flow around a cylinder with the Richardson number (Ri) of 1 over a duration of one second [5]. The quasi-steady nature of the flow allowed for the sufficient recovery of the fields using only 10 frames at different time steps, each frame comprising 200 data points for temperature and velocity fields obtained solely from boundary conditions. Utilizing a PINN architecture, composed of 10 hidden layers, each layer containing 100 neurons, and employing uniform weights for all loss functions, network optimization was conducted using the L-BFGS method. In another investigation, Cai et al. reconstructed a mixed convection flow within a 2D square domain, with heated bottom, operating at Ri = 1 [6]. Only temperature data was used to reconstruct the unknown pressure and velocity fields. Training dataset spanned from initial conditions to steady state, incorporating 100 frames, with each frame containing 9950 temperature data points. They emphasized the importance of establishing an appropriate weight ratio between the data and PDE losses to ensure convergence and prevent overfitting, recommending a weight ratio of 50:1 [6]. Lastly, in the same work [6], the demonstration of PINN for field reconstruction extends to experimental natural convection data. In this study, the unsteady three-dimensional (3D) temperature field in the flow over an espresso cup was captured using Tomographic Background Oriented Schlieren. Training focused on the transient phase, using 400 frames of temperature data without incorporating other boundary conditions. Employing a PINN architecture characterized by 10 hidden layers and 150 neurons per layer, the weight ratio between the data and PDE losses was set at 100:1.

Despite the successful demonstrations of PINNs for addressing inverse problems (i.e., field reconstructions) related to transient convection phenomena, a critical gap remains in understanding how the solution space of PDEs - all possible solutions under incompletely defined problem conditions - in the time domain influences PINN accuracy. Previous studies have primarily focused on either quasi-steady or unsteady states where temporal variations in the field, or equivalently, the time-derivatives of field variables, are not significant. Therefore, future research efforts may be directed towards adapting PINNs to accommodate varying levels of time dependence, which would require adjustments in the temporal domain of training data.

This work studies the impact of transience on the accuracy of PINNs in addressing unsteady convection problems. Specifically, the objective is to reconstruct the transient natural convection dynamics within a 2D square domain. Due to the increased complexity of the problem with significantly broadened solution space (i.e. possible field distributions along time domain) compared to previous research, we limit the spatial domain of the demonstration to 2D. The PINN is trained to infer completely unknown pressure ($P$), $x$- and $y$-velocity fields ($U$, $V$) from temperature data ($T$) available within an arbitrarily chosen time window without the knowledge of the prior flow field development.

A crucial consideration is the varying solution space of PDEs in the time domain across different temporal phases, which may not be uniformly captured by a PINN optimized under a single setting. For example, the incipient phase, characterized by large time derivatives, contrasts with the quasi-steady phase, where time derivatives approach zero. Addressing different temporal phases within transient processes may require adaptation of PINNs. Additionally, in a 2D natural convection phenomenon, multiple fields can exhibit variations across several spatial scales, leading to the vanishing gradient problem in PINNs. The efficacy of using a multiple scale method to mitigate the vanishing gradient problem is investigated.

## 2. Problem and methodology

### 2.1. Problem definition

The problem of interest is the reconstruction of the unknown field variables that are physically coupled: pressure, $x$- and $y$-velocity in a canonical transient buoyancy-driven flow within a 2D square cavity as shown in Fig. 1. The fields are reconstructed during an arbitrarily chosen time window if temperature data across the domain is provided, along with boundary conditions. Both width ($w$) and height ($h$) of the cavity are 0.015 m. The left wall is at $T_1 = 37°C$ and the right wall is at $T_0 = 35°C$. The top and bottom walls are thermally insulated (i.e., $dT/dx$ or

d$T$/d$y$ = 0), and all walls exhibit the no-slip condition. Asymmetric wall temperatures create a density gradient that drives a buoyancy-driven flow circulating the cavity. The flow field within the square cavity is characterized by the Rayleigh number, [Ra = $g\beta(T_{max} - T_{min})w^3/\nu\alpha$], where $g$ is the gravitational acceleration, $\beta$ is the thermal expansion coefficient, $\nu$ is the kinematic viscosity, and $\alpha$ is the thermal diffusivity. The temperature difference ($T_{max} - T_{min}$) is one way of controlling the complexity of flow fields, where $T_{max}$ is the maximum surface temperature, and $T_{min}$ is the minimum surface temperature. In this study, a temperature difference of 2°C results in a Ra of $2\times10^5$, which is below the critical Ra of $3\times10^5$, indicating that the flow remains laminar.

## 2.2. Governing equations

The 2D incompressible Navier-Stokes equation and the energy equation govern buoyancy-driven flow within a square domain. Non-dimensionalization of PDEs is often preferred to make the network adaptable to a wider array of problems by utilizing dimensionless terms to characterize the system. Furthermore, due to the nature of common activation functions used in PINNs (e.g. tanh and swish [2]), which exhibit nonlinearity and well-defined derivatives within the range of -1 to 1, normalizing the input and outputs within this interval is recommended. Consequently, the problem formulation incorporates the following normalized terms, denoted by '*' to signify nondimensional parameters with $x^* = x/w$, $y^* = y/h$, $t^* = t/t_{exp}$, $U^* = U/U_0$, $V^* = V/U_0$, $P^* = P/\rho_0 U_0^2$, $T^* = (T-T_{min})/(T_{max} - T_{min})$. The dimensional variables are defined as $x \in [0, w]$, $y \in [0, h]$, and $t \in [0, t_{exp}]$, where $t$ is the temporal coordinate and $t_{exp}$ is the experimental time duration, which defines the window of the reconstruction dataset. The characteristic variables are the reference velocity ($U_0$), maximum density ($\rho_0$), and reference temperature difference ($T_1 - T_0$). In order for Ri to be 1, $U_0$ was chosen to be $U_0 = [g\beta(T_1 - T_0)w]^{0.5}$, leading to $U_0 = 0.00992$ m/s. The Boussinesq approximation is used to capture density variation as a function of temperature and yields the following normalized PDEs.

$$\frac{\partial U^*}{\partial x^*} + \gamma \frac{\partial V^*}{\partial y^*} = 0 \quad (1)$$

$$St\frac{\partial U^*}{\partial t^*} + U^*\frac{\partial U^*}{\partial x^*} + \gamma V^*\frac{\partial U^*}{\partial y^*} + \frac{\partial P^*}{\partial x^*} - \frac{1}{Re}\left(\frac{\partial^2 U^*}{\partial x^{*2}} + \gamma^2 \frac{\partial^2 U^*}{\partial y^{*2}}\right) = 0 \quad (2)$$

$$St\frac{\partial V^*}{\partial t^*} + U^*\frac{\partial V^*}{\partial x^*} + \gamma V^*\frac{\partial V^*}{\partial y^*} + \gamma\frac{\partial P^*}{\partial y^*} - \frac{1}{Re}\left(\frac{\partial^2 V^*}{\partial x^{*2}} + \gamma^2 \frac{\partial^2 V^*}{\partial y^{*2}}\right) + RiT^* = 0 \quad (3)$$

$$St\frac{\partial T^*}{\partial t^*} + U^*\frac{\partial T^*}{\partial x^*} + \gamma V^*\frac{\partial T^*}{\partial y^*} - \frac{1}{Pe}\left(\frac{\partial^2 T^*}{\partial x^{*2}} + \gamma^2 \frac{\partial^2 T^*}{\partial y^{*2}}\right) = 0 \quad (4)$$

The dimensionless parameters St = $w/U_0 t_{exp}$, Re = $U_0 w/\nu$, Pe = RePr, and $\gamma$ represent the Strouhal number, Reynolds number, Peclet number, and geometric aspect ratio of the domain, respectively. According to the problem definition in section 2.1, these parameters are calculated as Re = 206, Pe = 997, and $\gamma$ = 1. We arbitrarily set $t_{exp}$ as 3.9 seconds that covers about 2.5 characteristic time periods ($t_c = w/U_0$), resulting in St of 0.39.

**2.3. Network architecture**

The PINN architecture used for the flow field reconstruction comprises a fully-connected neural network with 10 hidden layers, each housing 150 neurons (equivalently, 10×150 configuration). The select activation function is a locally adaptive sine function [3,7]. The governing equations demand an activation function that is second order differentiable and does not vanish to zero. The sine function has demonstrated efficacy in previous implementations when solving natural convection and mixed convection PDEs [8,9].

To explore hyperparameter sensitivity, the architecture was varied in both size and activation function. While a larger network offers increased expressiveness, it comes with the trade-off of higher computational cost [10]. Comparative tests were conducted using networks of different size (i.e. 12 hidden layers with 200 neurons per layer), yielding negligible differences, typically less than 1% when compared to networks of size 10×150. Smaller network size was not considered, since a previous work [6] reported large errors. Thus, the configuration of 10×150 was chosen. Additionally, an alternative activation function, 'tanh', was evaluated, in which no significant improvements were observed, thus the sine activation function was kept.

**2.4. Loss functions**

The loss functions used in PINNs constrain the reconstruction results to specified boundary conditions, data, and governing equations. Thus, the composite loss function ($L$) includes three components as follows:

$$L = L_{BC} + L_D + L_R \quad (5)$$

where $L_{BC}$ is the loss computed at points along the domain boundary, and $L_D$ is the loss evaluated at points where data is prescribed, denoted as data points. $L_R$ is the PDE loss calculated using both data points and additional points defined within the domain, referred to as collocation points. The losses are expressed as,

$$L_{BC} = \frac{1}{N_{BC}} \sum_{i=1}^{N_{BC}} |F(x_i, y_i, t_i) - F_i|^2 \quad (6)$$

$$L_D = \frac{1}{N_D} \sum_{i=1}^{N_D} |T^*(x_i, y_i, t_i) - T_i^*|^2 \quad (7)$$

$$L_R = \frac{1}{N_R + N_D} \sum_{i=1}^{N_R + N_D} \sum_{k=1}^{4} |e_k(x_i, y_i, t_i)|^2 \quad (8)$$

where $N_{BC}$, $N_D$, and $N_R$ represent the number of boundary points, data points, and collocation points, respectively. For $L_{BC}$, the field variables, $F$, reconstructed at the boundary are compared with the boundary conditions $F_i$. For $L_D$, reconstructed temperature values are compared against those obtained from CFD simulations, denoted as $T_i^*$. For $L_R$, the residuals between the reconstruction results and Eqns. (1 - 4), denoted as $e_1$ through $e_4$, are computed. The mean square error is used as the optimization criterion for PINNs.

In the composite loss, each loss term can be assigned a weight to bias the reconstruction towards a specific constraint. Previous studies have favored data over PDEs in PINNs by applying an $L_D/L_R$ weight ratio of 50-100 [6]. However, in our experiments, using an $L_D/L_R$ weight ratio from 1 to 100 did not yield substantial improvements. While a large $L_D/L_R$ weight ratio can accelerate convergence by aligning network parameters more quickly with data constraints, it may also compromise the generalization capacity of the PINNs. With a large $L_D/L_R$ weight ratio, the network parameters may deviate from accurately satisfying the PDEs, potentially leading to overfitting. Thus, equal weights were assigned to the loss terms for further analysis.

**2.5. Network optimization**

PINN was built using Pytorch libraries and was executed on a single Nvidia A100 GPU. The network optimization utilized floating-point precision with a machine epsilon of $1.19 \times 10^{-7}$. The optimization of the loss function employed a coupled strategy [11], initially utilizing the Adam optimizer with 100,000 epochs and a learning rate of 0.0001 followed by refinement with the L-BFGS optimizer until convergence [12,13]. The network parameters were initialized using the Xavier normal distribution.

To evaluate the reconstruction results and provide temperature data to the PINNs, reference data was generated using a finite volume model (FVM), implemented in ANSYS Fluent. The transient natural convection was simulated over a 60-second duration with a time step of 0.001 seconds, while data were extracted at intervals of 0.1 seconds (denoted as $\Delta t_{Data}$), resulting in a total dataset of 600 frames. From the total dataset, seven different subsets were sampled at distinct temporal phases of the flow, each spanning 3.9 seconds (equivalent to $t_{exp}$ and 40 frames). Table 1 lists the start and end times of the subsets. Although the start times $t_0$ differ among subsets, when the subset data is provided to a PINN, $t_0$ is normalized to $t^* = 0$, as illustrated in Fig. 1. The normalization of $t_0$ ensures that no information regarding the prior flow conditions or the absolute time ($t$) is provided to the PINN. This approach allows the PINN to operate independently of the absolute time of dataset, enabling flow field reconstruction at any point in time without relying on prior states.

Figure 2 illustrates a schematic overview of the field reconstruction process, displaying the points where the losses are computed. The number of boundary points in a single edge along the $x$, $y$, and $t$ axes, denoted as $N_x$, $N_y$, and $N_t$, is set to 100 each, resulting in a total of 40,000 boundary points, estimated as $N_b = [2(N_x+N_y)N_t]$. Consequently, the time step between boundary points along the $t$ axis ($\Delta t_{BC} = t_{exp}/N_t$) is 0.039 seconds. Collocation points, ranging from $N_r = 162,880$ to 262,880, are randomly sampled throughout the domain using the Latin-Hypercube (LHS) method [14]. Data points are also randomly sampled using LHS from the nodal information of the FVM data, representing 2.5% to 10% of the dataset (equivalently, between 893 to 3,572 points per frame). The coordinates of all points are provided to the PINN, where the loss is computed, and the network is updated through iterations.

## 2.6. Multiple scale method

Many thermofluidic systems exhibit flow features across different time or spatial scales. For instance, the spatial variations in pressure, velocity and temperature can differ significantly due to the mismatch in the diffusion coefficients of momentum and heat. In such multi-scale systems, we find that training the original PINN for all types of fields is challenging. To address these multi-scale issues, multiple scale analysis was performed.

Equations 1 through 4 are rescaled using the multiple scale method in perturbation theory. The PINN based on multiple scale method (MSPINN) closely aligns with previous works [15–18]. Huang et al. used *matched asymptotic expansions* of multiple networks to create a uniformly valid solution of governing equations with different scales [19]. Unlike Huang's approach, which solves outer and inner solutions separately using *matched asymptotic expansions*, we introduce augmented inputs via scaling factors directly integrated into the input layer, allowing the solution to be calculated with a single network. For Eqs. (1-4), we define the lengths ($\delta_{l,i}$) and time ($\delta_t$) scaling factors and corresponding rescaled variables, where $a_i = x^*/\delta_{l,i}$, $b_i = y^*/\delta_{l,i}$, and $c = t^*/\delta_t$, where the subscript $i$ denotes the $i^{th}$ scale. Thus, in MSPINN, the original PINN inputs, $x = [x^*, y^*, t^*]$, are expanded into a modified input structure, $\tilde{x} = [x^*, y^*, t^*, a_1, b_1, a_2, b_2, c]$. The modified input occupies a distinct layer following the initial network inputs $x$, as shown in Fig. 2(b).

To determine the scaling factors, the rescaled forms of Eqs. (1-4) are derived using the chain rule. For example, $\partial U^*/\partial x^* = (\partial U^*/\partial a_i)(\partial a_i/\partial x^*) + (\partial U^*/\partial x^*)(\partial x^*/\partial x^*) = U^*_{a,i}/\delta_{l,i} + U^*_x$, where the subscripts denote partial derivatives with respect to the corresponding variables. Most terms scale with unity, except for dimensionless numbers such as St, 1/Re and 1/Pe. Next, the magnitudes of each term in the rescaled Eqs. (1-4) are compared to identify the dominant terms. In Eq. (1), all terms are of the same scale, thus do not require further scale analysis. The scale analysis for Eqs. (2) and (3) is identical due to their structural similarity and the fact that Ri = 1. The rescaled forms of Eqs. (2) and (4) are presented as follows:

$$St(U^*_{c_1}\frac{1}{\delta_{t,1}} + U^*_t) + U^*(U^*_{a_1}\frac{1}{\delta_{l,1}} + U^*_x) + \gamma V^*(U^*_{b_1}\frac{1}{\delta_{l,1}} + U^*_y) + (P^*_{a_1}\frac{1}{\delta_{l,1}} + P^*_x) -$$

$$\frac{1}{Re}\left[(U^*_{a_1 a_1}\frac{1}{\delta_{l,1}^2} + \frac{2}{\delta_{l,1}}U^*_{a_1 x} + U^*_{xx}) + \gamma^2(U^*_{b_1 b_1}\frac{1}{\delta_{l,1}^2} + \frac{2}{\delta_{l,1}}U^*_{b_1 y} + U^*_{yy})\right] = 0 \qquad (9)$$

$$St(T^*_{c_2}\frac{1}{\delta_{t,2}} + T^*_t) + U^*(T^*_{a_2}\frac{1}{\delta_{l,2}} + T^*_x) + \gamma V^*(T^*_{b_2}\frac{1}{\delta_{l,2}} + T^*_y) -$$

$$\frac{1}{Pe}\left[(T^*_{a_2 a_2}\frac{1}{\delta_{l,2}^2} + \frac{2}{\delta_{l,2}}T^*_{a_2 x} + T^*_{xx}) + \gamma^2(T^*_{b_2 b_2}\frac{1}{\delta_{l,2}^2} + \frac{2}{\delta_{l,2}}T^*_{b_2 y} + T^*_{yy})\right] = 0 \quad (10)$$

In Eq. (9), the order of magnitudes for each term is represented by the coefficients of $T^*$, $U^*$, $V^*$, $P^*$ terms, such as $St/\delta_{t,1}$, $St$, $1/\delta_{l,1}$, 1, $1/\delta_{l,1}^2 Re$, $1/\delta_{l,1} Re$, and $1/Re$. For simplicity in analysis, we assume that the velocities $U^*$ and $V^*$ are of the same order of magnitude as $1/\delta_{l,1}$. Consequently, the terms $1/\delta_{l,1}^2 Re$ and $St/\delta_{t,1}$ are considered dominant. By balancing the dominant terms with the constant term (i.e., 1), we obtain $\delta_{l,1} = Re^{-0.5}$ and $\delta_{t,1} = St$. Similarly, the scale analysis of Eq. (10) leads to a dominant balance, yielding $\delta_{l,2} = Pe^{-0.5}$ and $\delta_{t,2} = St$. Holmes provides in-depth explanations on dominant balance and multiple-scale analysis [20]. Note that these scaling factors should be updated for different problems, as each problem will have its own unique scaling relations. However, problems governed by the same PDEs may share the same scaling constants, thus eliminating the need for repeated derivation.

## 3. Results and discussion

### 3.1. Optimization of traditional PINNs

To assess the efficacy of traditional PINNs, baseline PINNs without rescaled input variables (PINN$_0$) were first optimized using the most resource-intensive settings. These settings utilized 10% of the available FVM data points (i.e., $N_d$ = 142,880 points), with $N_b$ = 40,000, and $N_r$ = 162,880. The network optimization time was influenced by factors such as the number of epochs and the volume of collocation points. On average, completing 100,000 epochs while processing $N_r$ of 162,880 took about six hours and used around 40 GB of memory.

Figure 3 compares the reconstruction results of PINN$_0$ at three different flow phases (i.e., sets 1, 5, 7) with the FVM reference data. The reconstructed fields show good qualitative agreement with the reference data. Initially, the warm left wall leads to an upward buoyancy-driven flow, represented by a large $V^*$, resulting in pressurization at the top left corner characterized by an elevated $P^*$. During the development phase, the pressurized, high-temperature fluid at the top left corner moves along the top wall towards the cold right wall, inducing a circulating flow pattern. Along the cold right wall, the fluid descends, facilitating the formation of stable convection cells within the cavity. For a visualization purpose, FVM maps were linearly interpolated to the same grid resolution of $PINN_0$ (i.e., 500×500).

Next, to identify an efficient network optimization setting, the traditional PINN was optimized using two datasets - set 2 or set 3 - under various configurations. These datasets represent the most transient flow regimes, characterized by the largest temperature gradients over time. The most transient regimes are typically recognized as more challenging for optimizing PINN. In this experiment, the mean absolute error, $\varepsilon$ (MAE), was used to quantitatively evaluate the performance of the PINN given as $\varepsilon=|F-\hat{F}|_1/N$, where $F$, $\hat{F}$, and $N$ are the PINN prediction, FVM ground truth, and number of points, respectively. Errors were calculated for each timestep across all grid nodes. Due to the stochastic nature of network optimization, errors varied in each optimization. Thus, the PINN underwent five optimization iterations.

Experiments revealed that increasing $N_r$ from 162,880 to 262,880 did not lead to significant improvements in errors, thus $N_r$ of 162,880 was chosen. The variation in $N_d$ from 2.5% to 10% of total data resulted in minor but noticeable differences. It was found that at least 5% of data should be used for the network optimization, and 10% was chosen to ensure the PINN accurately captures the detailed features of the flow. Tuning the learning rate showed that $1\times10^{-4}$ provided the optimal balance between error reduction and optimization time. For example, about 3% decrease in error was traded for a doubled optimization time. Lastly, it was determined that 100,000 epochs were optimal. Longer epochs are usually required for tackling more complex problems. The hyperparameters derived in this section are used for subsequent reconstructions. For detailed descriptions of these experiments, refer to the supplementary material.

### 3.2. Reconstructions by PINN and MSPINN

The transient evolution of the flow can be characterized into three distinct phases: the incipient phase ($R_{in}$), development phase ($R_{de}$) and quasi steady-state phase ($R_{ss}$), each exhibiting unique features. In the incipient phase [Fig. 3(a, b)], the temperature and velocity fields display relatively uniform distributions, with a significant solution space of the next phase and large field gradients with respect to time. In this context, the 'solution space' refers to the range of possible field distributions that the PINN can converge to. Next, the development phase [Fig. 3(c, d)] covers the majority of transient behaviors, with the fields having the largest time derivatives. Lastly, the quasi steady-state phase [Fig. 3(e, f)] is the transition from development to steady-state, in which the temporal effects become less apparent. The differences among three phases can impact the accuracy of PINN reconstruction, particularly considering the varying solution space.

To understand the impact of flow transience on PINN reconstruction, both the traditional PINN and MSPINN were used to reconstruct the fields at six flow phases (i.e., Set 1, 3, 4, 5, 6, and 7). Each reconstruction underwent three to five iterations of network optimization, with the subsequent analysis based on the average results of these iterations. Figure 4 shows $\varepsilon$ of both original PINNs and MSPINNs, offering several interesting observations. Firstly, errors are relatively large at the incipient phase ($R_{in}$) and diminish at the quasi-steady state ($R_{ss}$). Secondly, within each flow phase, there is wide variation in error. For a statistical analysis of $\varepsilon$, we estimate the mean (denoted as $\bar{\varepsilon}$) and range (denoted as $\Delta\varepsilon$), defined as the difference between maximum and minimum values of $\varepsilon$. Despite a few outliers, both $\bar{\varepsilon}$ and $\Delta\varepsilon$ decrease over time, indicating that PINNs are more accurate during $R_{ss}$ than during $R_{in}$.

The comparison between the original PINNs and MSPINNs show that improvements can be made by rescaling, in which the impact is more prominent in earlier velocity reconstructions. The maximum improvement in $\varepsilon$ was around $3.78 \times 10^{-2}$ in $U^*$ (roughly 72.2%), and on average the improvement in $\varepsilon$ across all field variables was around $1.96 \times 10^{-3}$ (around 6.4%). This demonstrates that MSPINN effectively accounts for the various scales present in the natural convection process. Additionally, Fig. 5 compares the fields reconstructed by PINN and MSPINN during a development phase (i.e., frame 140). While the PINN captures large-scale temperature patterns, the MSPINN provides more accurate reconstructions of sharp temperature gradients in localized regions.

### 3.3. Error sources

We introduce two metrics to understand the cause of relatively large inaccuracies observed during the incipient phase. The first metric, namely the temporal gradient ($\nabla_t \hat{F} = |\hat{F}_{t+1} - \hat{F}_t|/\Delta t$), is used to investigate the cause of relatively large inaccuracies in the incipient phase, where $\hat{F}$ is the field variable obtained from FVM groundtruth and $\Delta t$ is the time step. We hypothesize that the size of the solution space influences the network performance. This hypothesis stems from the observation that PINNs tend to exhibit higher errors in systems with a larger solution space, especially evident in ill-posed reconstruction problems. A greater solution space implies more possible independent ways in which the field variables can vary over time. Therefore, a higher temporal gradient (denoted as $\nabla_t \hat{F}$) is expected in systems with greater solution space. Accordingly, we indirectly characterize the size of solution space in the system by using $\nabla_t \hat{F}$.

Different solution spaces could exist under various timescales. For example, the incipient phase will include large temporal derivatives (equivalently larger solution space), while the quasi-steady phase will have near zero temporal derivatives (equivalently smaller solution space). These phases, each with significantly different solution space, may necessitate PINNs optimized under different settings.

Figure 6 presents $\nabla_t\hat{F}$, where $\nabla_t\hat{F}$ is an order of magnitude greater during the incipient phase than during the quasi-steady phase. Thus, the incipient phase is shown to possess a larger solution space. It is also clear that the relative magnitude of $\nabla_t\hat{F}$ loosely correlates to the magnitude of $\Delta\varepsilon$ and $\bar{\varepsilon}$, in which a larger $\nabla_t\hat{F}$ roughly corresponds to a higher $\Delta\varepsilon$ and $\bar{\varepsilon}$.

The second metric is the absolute spatial gradients ($\nabla_s\hat{F}=|\hat{F}_{s+1} - \hat{F}_s|/\Delta s$) of each field, where $\Delta s$ is the spatial step along either the $x$ ($\Delta s_x$) or $y$ ($\Delta s_y$) axis that in general can be different. Figure 7 depicts $|\nabla_s\hat{F}|$ on a logarithm base 10 scale. Three frames were considered, i.e., $10^{th}$, $200^{th}$, and $500^{th}$ frames, representing different phases of field development. To compute the gradients, FVM data was interpolated on a 500×500 grid with $\Delta s$ of $3\times10^{-5}$ m. Figure 7 shows that the $|\nabla_s\hat{F}|$ spans over four orders of magnitude, from $10^{-6}$ to $10^{-2}$. When $|\nabla_s\hat{F}|$ becomes very small, it results in small gradient terms in the loss function. This, in return, hinders effective parameter updates during backpropagation through deep neural networks, a phenomenon referred to as the vanishing gradient issue [4]. Figure 7 (c, f, i) also shows the distribution of errors in each field at $10^{th}$, $200^{th}$, and $500^{th}$ frames. The pressure errors exhibit relatively uniform distribution, while exhibiting large errors appear near the cold wall at the incipient phase where the $|\nabla_x P|$ is the smallest. For the temperature field, errors are less than $2\times10^{-2}$, since the PINN was constrained by the temperature data. For the velocity fields, errors are highest during the incipient phase and around the areas with near-zero gradients. The alignment of regions characterized by extremely small $|\nabla_s\hat{F}|$ and large $\varepsilon$ supports the presence of a vanishing gradient issue, particularly noticeable during the incipient phase.

Additionally, Fig. 7 reveals multiple length scales within the field. First, the spatial scale of the gradient field variation is relatively small. Gradient field values exhibit an order of magnitude variation even across $\Delta x^*$ or $\Delta y^*$ of 0.1. However, the spatial scale of original fields shown in Fig. 3 is relatively large. An order of magnitude variation of original field variables occurs across the length of the entire domain. If there are multiple length scales between the original and gradient fields, it necessitates rescaling the PDEs in Eqs. 1 - 4. Otherwise, PINNs tend to favor solutions preserving the larger-scale variations, while the small-scale variations become recessive due to the vanishing gradient issue.

## 4. Conclusions

This paper explored the ability of PINNs to reconstruct pressure and velocity fields from an instantaneous temperature data for a transient natural convection process. A fully connected network with a structure of 10 layers and 150 neurons per layer was employed to construct the PINN, and its performance was enhanced through hyperparameter tuning. Study into the effects of transience on PINNs led to a few key results. PINN exhibited relatively higher accuracies during the quasi-steady state, while less accuracies during the incipient state. To explain the inaccuracies during the incipient phase, we examined two metrics: spatial and temporal gradients of the field variables ($\nabla_s \hat{F}$ and $\nabla_t \hat{F}$). The temporal gradients of field variables serve as indicators of the solution space, complexity and variability within a field. As expected, magnitude of $\nabla_t \hat{F}$ was an order of magnitude greater during the incipient phase than during the quasi-steady phase, indicating the presence of significantly larger solution space during the incipient phase. The spatial gradients of field variables revealed that $\nabla_s \hat{F}$ spanned more than four orders of magnitude in the natural convection studied, even reaching down to $10^{-6}$. The correlation between regions exhibiting extremely small $|\nabla_s \hat{F}|$ and large $\varepsilon$ indicated the presence of a vanishing gradient problem, particularly evident during the incipient phase.

Furthermore, we noticed various length scales present within the field. To reconstruct the fields at different scales, we employed multiple scale method, in which the inputs were recast into three distinct scales. The MSPINN resulted in lower $\varepsilon$ with maximum and average improvements of around 72.2% and 6.4%, respectively. This result demonstrates the potential efficacy of the multiple scale method in solving thermofluidic reconstruction problems where solutions span across different scales. Though scaling is currently performed manually, research into perturbation theory could automate this process. Moreover, generalization of MSPINN to other systems and experimental data will further solidify the applicability of this method, including the turbulent systems which necessitate solving multiple scale equations.

**Acknowledgment**

This work was supported by two National Science Foundation grants under Grant No. 2053413 and 2337973. Additionally, this study was partially supported by Arizona State University startup funds.

**Table 1.** Seven datasets used for field reconstructions.

|                          | Set 1 | Set 2 | Set 3 | Set 4 | Set 5 | Set 6 | Set 7 |
|--------------------------|-------|-------|-------|-------|-------|-------|-------|
| Start frame              | 10    | 60    | 100   | 150   | 200   | 300   | 500   |
| Start time $t_0$ [s]     | 1     | 6     | 10    | 15    | 20    | 30    | 50    |
| End frame                | 49    | 99    | 139   | 189   | 239   | 339   | 539   |
| End time $t_0 + t_{exp}$ [s] | 4.9 | 9.9 | 13.9 | 18.9 | 23.9 | 33.9 | 53.9 |

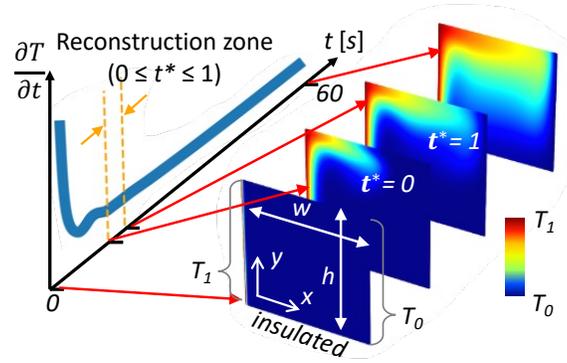

**Figure 1.** Schematic of a square cavity with a heated left wall, a cooled right wall, and insulated top and bottom walls. During the initial stage, the temperature field exhibits a significant temporal gradient. A reconstruction zone is arbitrarily selected within the complete dataset, and a normalized time scale ($0 \leq t^* \leq 1$) is assigned.

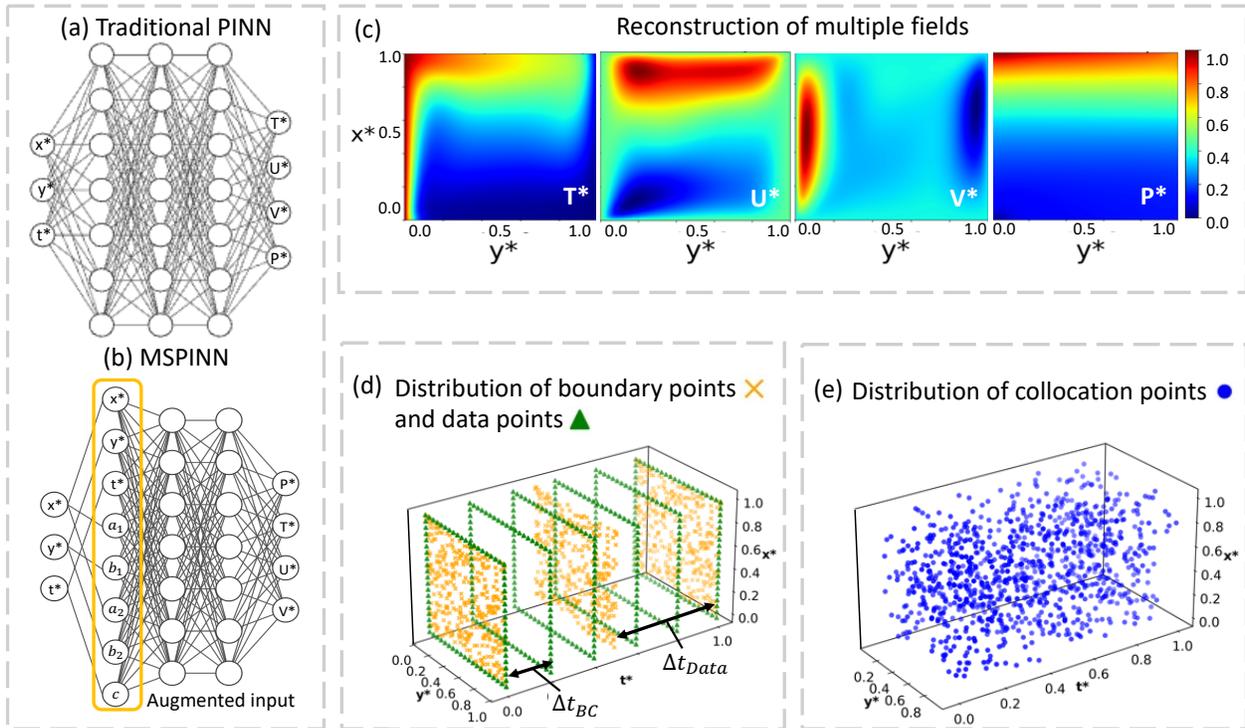

**Figure 2.** Overview of field reconstruction process. Two PINN architectures are tested: (a) a traditional PINN and (b) a PINN modified based on multiple scale method (MSPINN). PINNs reconstruct unknown fields ($U^*$, $V^*$, $P^*$) based on the provided temperature field ($T^*$). Neural network losses are evaluated at (d) boundary and data points, as well as at € collocation points.

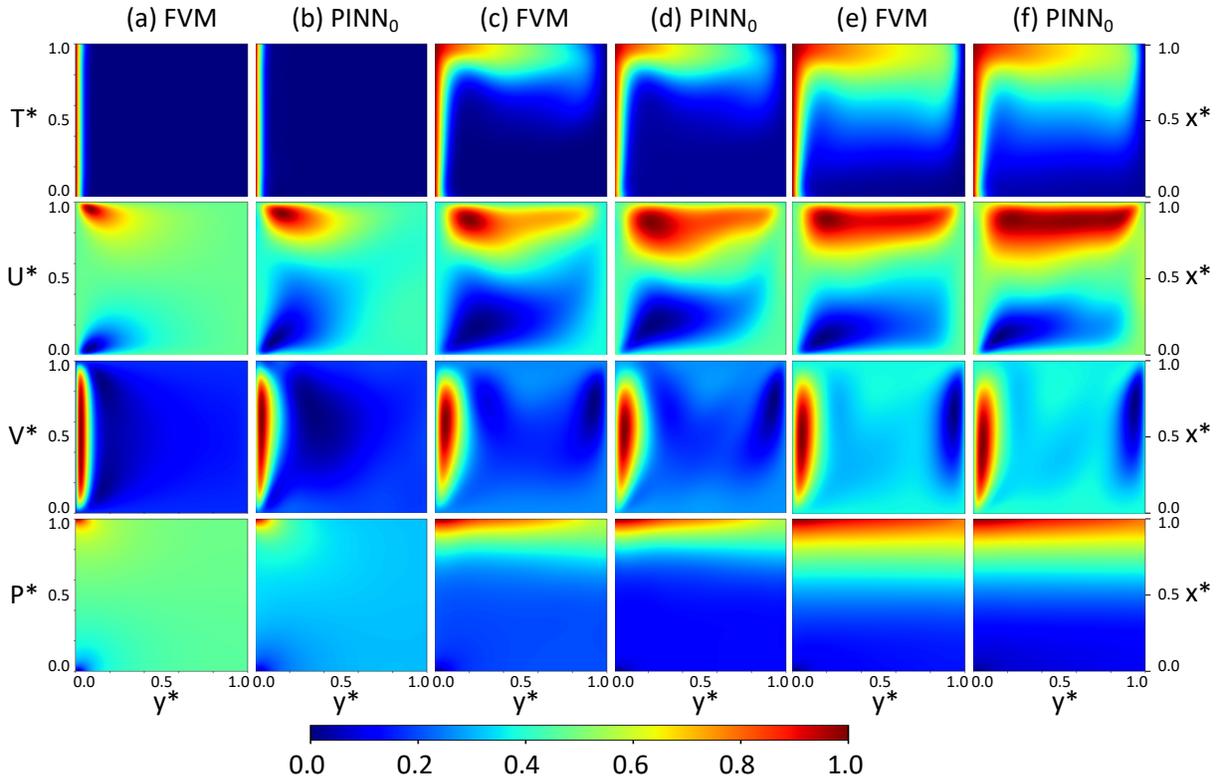

**Figure 3.** Comparison of FVM (a, c, e) and PINN$_0$ predictions (b, d, f), evaluated at three different flow phases: (a, b) Incipient phase at $t$ = 1s (start frame of set 1), (c, d) development phase at $t$ = 20s (start frame of set 5), and (e, f) quasi-steady state at $t$ = 50s (start frame of set 7).

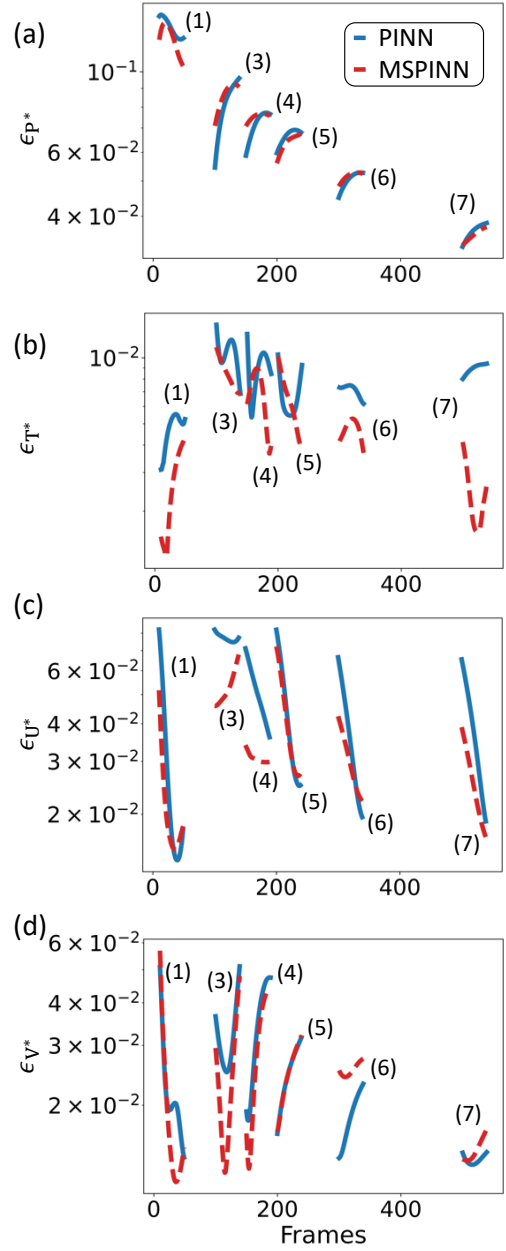

**Figure 4.** $\varepsilon$ of PINN and MSPINN for the six different temporal sets. Numbers in each plot represent the dataset number.

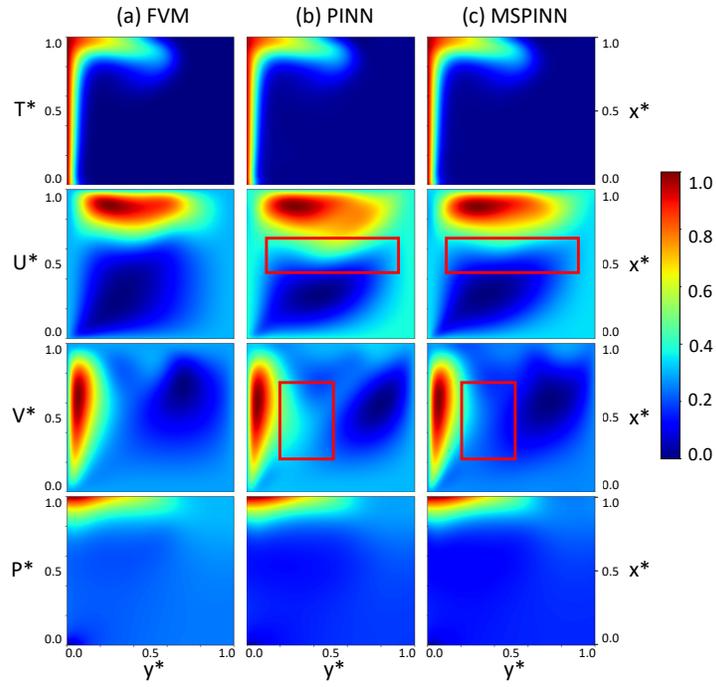

**Figure 5.** Comparison between FVM, PINN, and MSPINN at frame 140 of dataset. The improvements in reconstructions are most prominent at the upper surface of boundary layers and are indicated by the boxed regions.

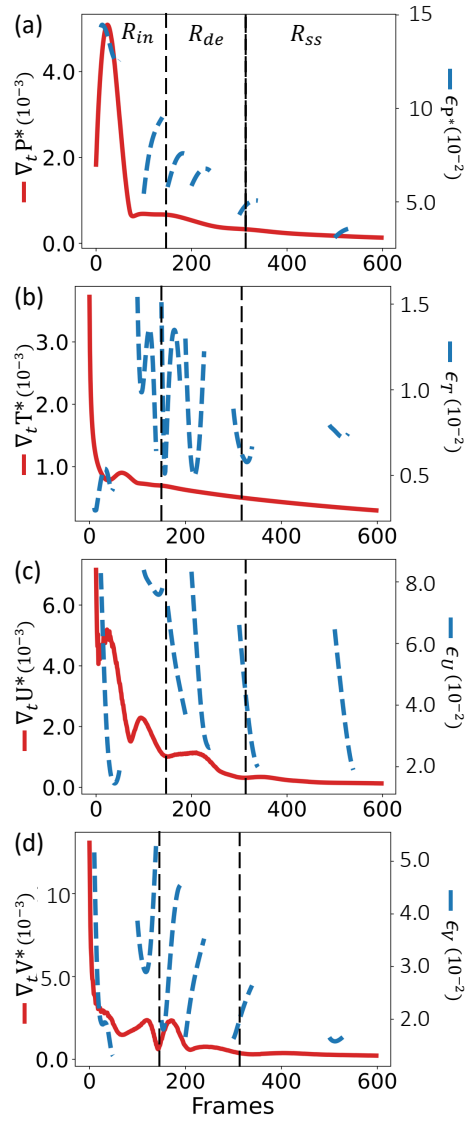

**Figure 6.** $\nabla_t \hat{F}$ for each field (solid) overlaid with $\varepsilon$ of PINN (dashed).

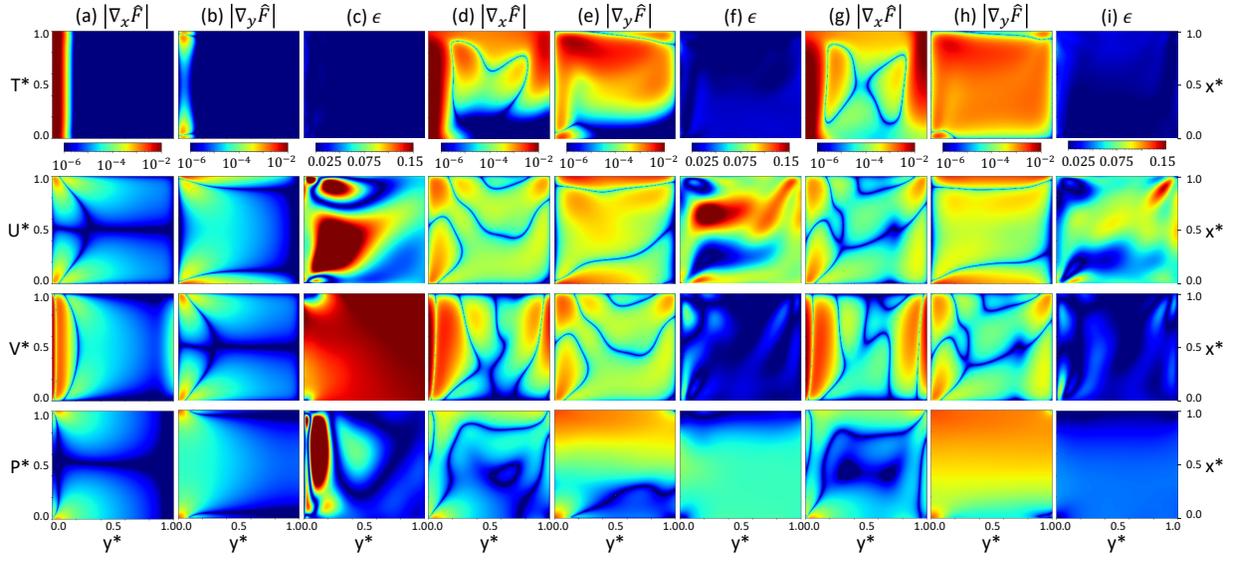

**Figure 7.** $|\nabla_s \hat{F}|$ for x- and y-axis, and $\varepsilon$ of field variables obtained at (a,b,c) $10^{th}$ frame, (d,e,f) $200^{th}$ frame, and (g,h,i) $500^{th}$ frame. The fields are plotted on a clipped log scale between $10^{-2}$ and $10^{-6}$ to enhance contrast. Error fields are shown with color scales clipped between $5\times10^{-3}$ and $1.5\times10^{-1}$ to show important contrast.

# References

[1] Raissi, M., Perdikaris, P., and Karniadakis, G. E., "Physics-Informed Neural Networks: A Deep Learning Framework for Solving Forward and Inverse Problems Involving Nonlinear Partial Differential Equations," *Journal of Computational Physics*, Vol. 378, 2019, pp. 686–707. https://doi.org/10.1016/j.jcp.2018.10.045

[2] Cuomo, S., di Cola, V. S., Giampaolo, F., Rozza, G., Raissi, M., and Piccialli, F., "Scientific Machine Learning through Physics-Informed Neural Networks: Where We Are and What's Next," arXiv, Jun2022. https://doi.org/10.48550/arXiv.2201.05624

[3] Cai, S., Wang, Z., Wang, S., Perdikaris, P., and Karniadakis, G. E., "Physics-Informed Neural Networks for Heat Transfer Problems," *Journal of Heat Transfer*, Vol. 143, No. 6, 2021, p. 060801. https://doi.org/10.1115/1.4050542

[4] Lucor, D., Agrawal, A., and Sergent, A., "Physics-Aware Deep Neural Networks for Surrogate Modeling of Turbulent Natural Convection," arXiv, Mar2021.

[5] Wang, T., Huang, Z., Sun, Z., and Xi, G., "Reconstruction of Natural Convection within an Enclosure Using Deep Neural Network," *International Journal of Heat and Mass Transfer*, Vol. 164, 2021, p. 120626. https://doi.org/10.1016/j.ijheatmasstransfer.2020.120626

[6] Cai, S., Wang, Z., Fuest, F., Jeon, Y.-J., Gray, C., and Karniadakis, G. E., "Flow over an Espresso Cup: Inferring 3D Velocity and Pressure Fields from Tomographic Background Oriented Schlieren Videos via Physics-Informed Neural Networks," *Journal of Fluid Mechanics*, Vol. 915, 2021, p. A102. https://doi.org/10.1017/jfm.2021.135

[7] Jagtap, A. D., Kawaguchi, K., and Em Karniadakis, G., "Locally Adaptive Activation Functions with Slope Recovery for Deep and Physics-Informed Neural Networks," *Proceedings of the Royal Society A: Mathematical, Physical and Engineering Sciences*, Vol. 476, No. 2239, 2020, p. 20200334. https://doi.org/10.1098/rspa.2020.0334

[8] Raissi, M., Wang, Z., Triantafyllou, M. S., and Karniadakis, G. E., "Deep Learning of Vortex Induced Vibrations," *Journal of Fluid Mechanics*, Vol. 861, 2019, pp. 119–137. https://doi.org/10.1017/jfm.2018.872

[9] Raissi, M., Yazdani, A., and Karniadakis, G. E., "Hidden Fluid Mechanics: Learning Velocity and Pressure Fields from Flow Visualizations," *Science*, Vol. 367, No. 6481, 2020, pp. 1026–1030. https://doi.org/10.1126/science.aaw4741

[10] Raghu, M., Poole, B., Kleinberg, J., Ganguli, S., and Sohl-Dickstein, J., "On the Expressive Power of Deep Neural Networks," arXiv, Jun2017.

[11] Eivazi, H., Tahani, M., Schlatter, P., and Vinuesa, R., "Physics-Informed Neural Networks for Solving Reynolds-Averaged Navier–Stokes Equations," *Physics of Fluids*, Vol. 34, No. 7, 2022, p. 075117. https://doi.org/10.1063/5.0095270

[12] Kingma, D. P., and Ba, J., "Adam: A Method for Stochastic Optimization," arXiv, Jan2017.

[13] Liu, D. C., and Nocedal, J., "On the Limited Memory BFGS Method for Large Scale Optimization," *Mathematical Programming*, Vol. 45, Nos. 1–3, 1989, pp. 503–528. https://doi.org/10.1007/BF01589116

[14] "Randomized Designs — pyDOE 0.3.6 Documentation."

[15] Callaham, J. L., Koch, J. V., Brunton, B. W., Kutz, J. N., and Brunton, S. L., "Learning Dominant Physical Processes with Data-Driven Balance Models," *Nature*

*Communications*, Vol. 12, No. 1, 2021, p. 1016. https://doi.org/10.1038/s41467-021-21331-z

[16] Weng, Y., and Zhou, D., "Multiscale Physics-Informed Neural Networks for Stiff Chemical Kinetics," *The Journal of Physical Chemistry A*, Vol. 126, No. 45, 2022, pp. 8534–8543. https://doi.org/10.1021/acs.jpca.2c06513

[17] Wang, S., Wang, H., and Perdikaris, P., "On the Eigenvector Bias of Fourier Feature Networks: From Regression to Solving Multi-Scale PDEs with Physics-Informed Neural Networks," *Computer Methods in Applied Mechanics and Engineering*, Vol. 384, 2021, p. 113938. https://doi.org/10.1016/j.cma.2021.113938

[18] Zhang, L., and He, G., "Multi-Scale-Matching Neural Networks for Thin Plate Bending Problem," *Theoretical and Applied Mechanics Letters*, Vol. 14, No. 1, 2024, p. 100494. https://doi.org/10.1016/j.taml.2024.100494

[19] Huang, J., Qiu, R., Wang, J., and Wang, Y., "Multi-Scale Physics-Informed Neural Networks for Solving High Reynolds Number Boundary Layer Flows Based on Matched Asymptotic Expansions," *Theoretical and Applied Mechanics Letters*, Vol. 14, No. 2, 2024, p. 100496. https://doi.org/10.1016/j.taml.2024.100496

[20] Holmes, M. H., "Introduction to Perturbation Methods," Springer Science & Business Media, 2012.

[21] Jin, X., Cai, S., Li, H., and Karniadakis, G. E., "NSFnets (Navier-Stokes Flow Nets): Physics-Informed Neural Networks for the Incompressible Navier-Stokes Equations," *Journal of Computational Physics*, Vol. 426, 2021, p. 109951. https://doi.org/10.1016/j.jcp.2020.109951

# Supplementary Material

A. Network tuning

    A.1. Effect of network randomization

    A.2. Effect of collocation points

    A.3. Effect of data points

    A.4. Effect of Adam learning rate

    A.5. Effect of epochs

## A. Network tuning

### A.1. Effect of network randomization

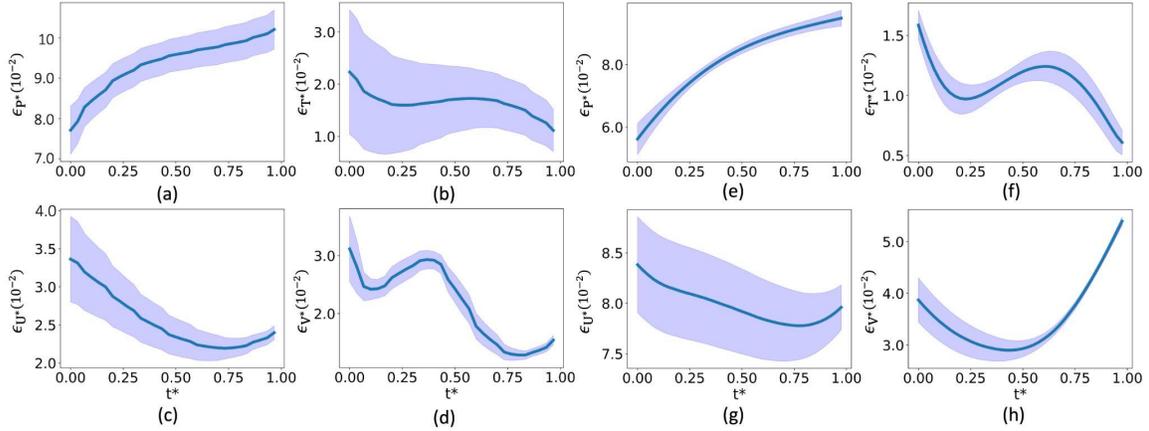

Figure A.1. $\varepsilon$ of $PINN_0$ obtained with set 2 (incipient phase, a-d) and set 3 (intermediate phase, e-h).

The standard deviation of $\varepsilon$ represents the variation due to randomness within the PINN, serving as a metric to identify the significance of varying hyperparameters. Noticeable standard deviations in temperature and x-velocity (Fig. A.1b, A.1c) are apparent for set 2. Similarly, large standard deviations in x-velocity prediction are observed for set 3 (Fig. A.1g).

After establishing the training results of $PINN_0$ under the conservative setting, the training setting of $PINN_0$ was modified using set 2 to optimize the number of collocation points ($N_r$), number of data points ($N_d$), and learning rate of the Adam optimizer, followed by using set 3 to optimize the number of epochs. At each hyperparameter setting, three training iterations were carried out, and the best iteration was determined through visual inspection. Then, $\varepsilon$ values of the best iteration were employed to study the effects of hyperparameters.

## A.2. Effect of collocation points

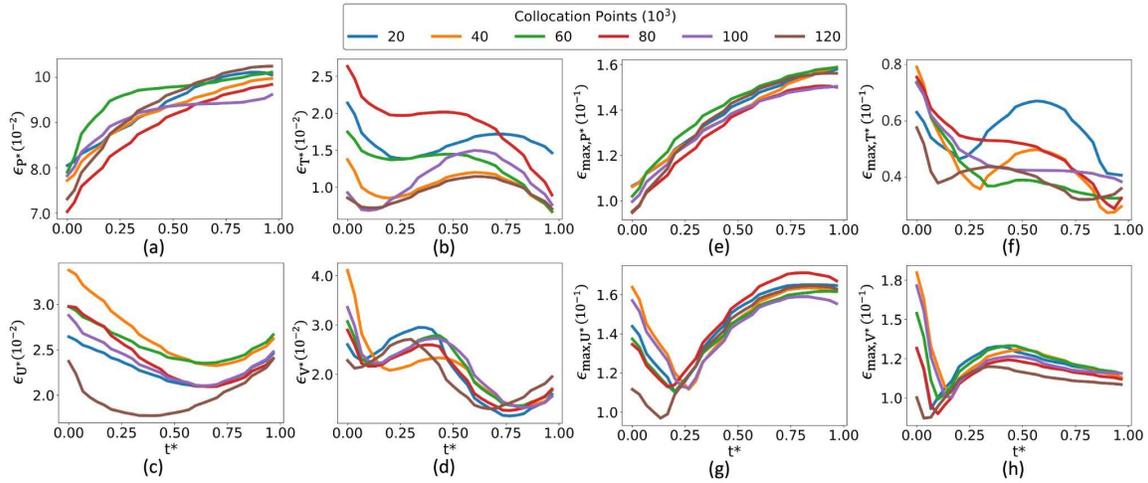

Figure A.2: (a-d) $\varepsilon$ and (e-h) $\varepsilon_{max}$ for different collocation points. Minimal improvements are seen between Nr: 20,000 and 120,000.

The number of collocation points determines the extent to which the network enforces the governing equations to its solutions, which is analogous to the refinement of mesh grids in CFD. Using set 2, experimentation involved varying Nr: 20, 40, 60, 80, 100, and 120 thousand. Note that all data point nodes also functioned as additional collocation points. Thus, the total number of points used to calculate the PDE loss is the sum of data points Nd and Nr (e.g. 20,000 of Nr + 142,880 of Nd = 162,880 of total Nr). The results of varying Nr are shown in Fig. A.2. The impact of Nr on the variation of $\varepsilon$ was pronounced for temperature and x-velocity, with $\varepsilon$ difference between the best and worst cases being about $1.5 \times 10^{-2}$ and $1 \times 10^{-2}$, respectively. Generally, an increase of Nr led to a slight reduction in $\varepsilon$ for temperature and x-velocity. However, this improvement falls within the expected range of variation and is considered insignificant. Thus, employing Nr of 20,000 is deemed sufficient for subsequent analyses.

## A.3. Effect of data points

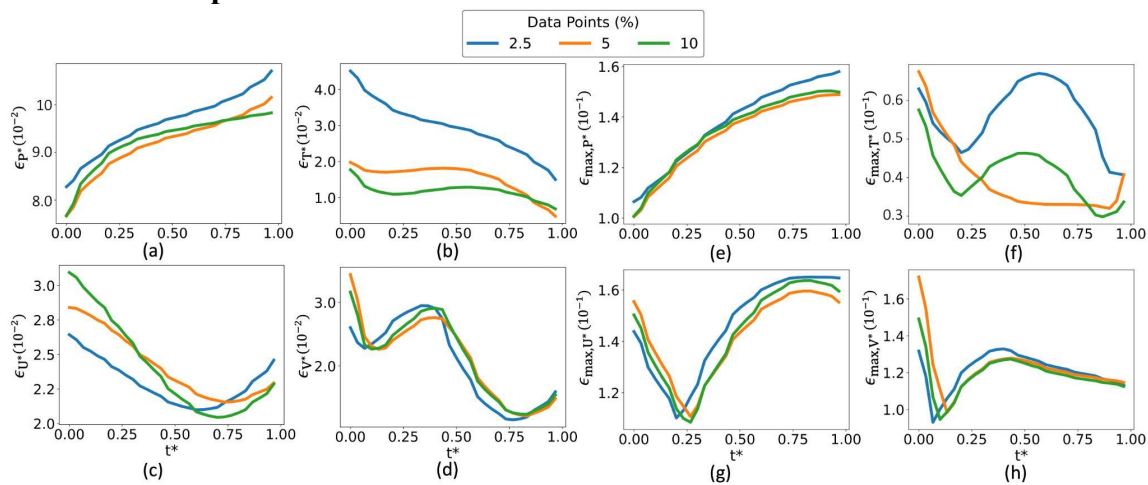

Figure A.3. (a-d) $\varepsilon$ and (e-h) $\varepsilon_{max}$ for different data points. A minimum of 5% is found to be necessary.

Similarly, the number of data points Nd determines the extent to which the network enforces temperature constraints to its solutions. Since the PINN uses temperature data to reconstruct other fields, a larger Nd can enhance solution uniqueness. Using set 2, experimentation involved 2.5%, 5%, and 10% of the entire dataset, corresponding to 893, 1786, and 3572 of Nd per frame, respectively. Across all trials, a total Nr of 162,880 were used. To maintain a consistent contribution from the PDE loss Lr, the total Nd was subtracted from the total Nr (162,880) to calculate the additional Nr [e.g. 162,880 of total Nr = (890 Nd/frame) (40 frames) + 127,280 of extra Nr]. The results of varying Nd are shown in Fig. A.3. The impact of Nd on the variation of $\varepsilon$ was pronounced for temperature, with $\varepsilon$ difference between the best and worst cases being about $2.8\times10^{-2}$. Interestingly, the decrease in temperature accuracy due to fewer Nd did not significantly impact the reconstruction of other fields. Nevertheless, lower errors suggest that at least 5% of the data to be used for further analyses.

## A.4. Effect of Adam learning rate

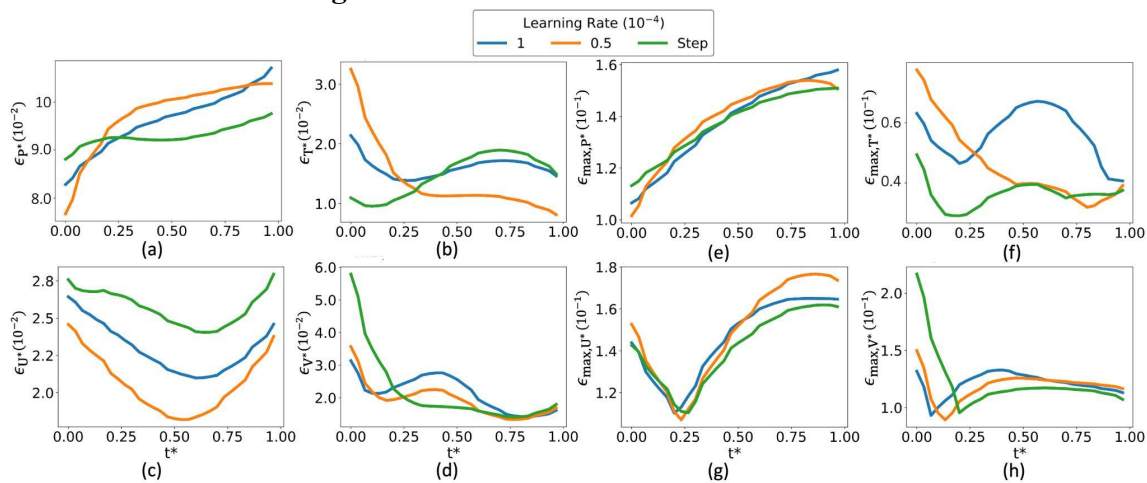

Figure A.4. (a-d) $\varepsilon$ and (e-h) $\varepsilon_{max}$ for Adam learning rates. The smaller rate generally yielded less error. Interestingly, the step learning rate underperformed compared to the fixed rate.

The learning rate of Adam optimizer determines how fast the network steps through the gradient descent process. A large rate may cause overshooting of optima, leading to network instability, while too small rate can result in slow convergence. To mitigate these issues, schedulers are often employed to incrementally decrease the learning rate based on certain criteria such as loss plateau or number of epochs. Here, set 2 was used, and three different learning rates were experimented, namely, $1\times10^{-4}$, $5\times10^{-5}$ and a scheduler scheme. Each model was trained for 100,000, 200,000, and 200,000 Adam epochs, respectively. The scheduler scheme consists of 100,000 epochs at a rate of $1\times10^{-4}$, followed by an additional 100,000 epochs at a rate of $5\times10^{-5}$. Subsequently, all learning beyond the Adam epochs underwent refinement with L-BFGS to yield the final results. The results of varying learning rates are shown in Fig. A.4. Predictions for temperature and y-velocity were more impacted than the other variables, with $\varepsilon$ difference between the best and worst cases being about $2.3\times10^{-2}$ and $2.7\times10^{-2}$, respectively. Despite the observed deviation in temperature being within the expected range, the y-velocity plot indicates better convergence with a single learning rate. Additionally, x-velocity predictions showed slight improvements with a smaller learning rate and without the step scheme. The tradeoff between slight accuracy improvement (10% decrease in $\varepsilon$) and time (50% increase in compute time) suggests that a learning rate of $1\times10^{-4}$ suffices for the reconstruction tasks.

## A.5. Effect of epochs

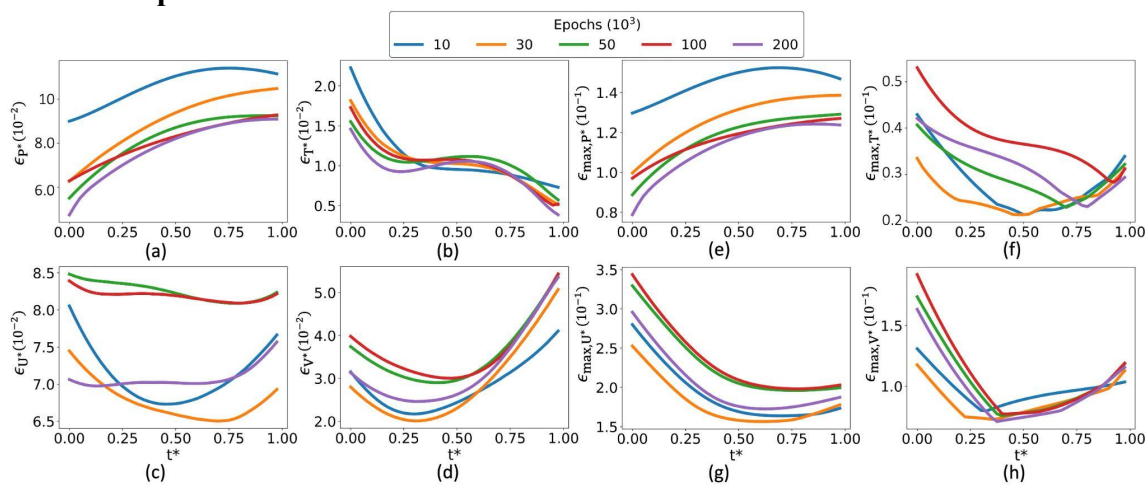

Figure A.5. (a-d) $\varepsilon$ and (e-h) $\varepsilon_{max}$ for different Adam training epochs. Longer epochs generally yielded less error. Interestingly, epochs in the middle range (50,000 and 100,000) had larger errors compared to the small range (10,000 and 30,000) and largest epoch (200,000).

The number of Adam epochs determine the initialization for the L-BFGS optimizer. Typically, a larger number of Adam epochs results in a lower initial loss for L-BFGS. This coupled method leverages the strengths of each optimizer [11,21]. The Adam optimizer, a well-implemented stochastic gradient descent (SGD) method, is used to initialize the network close to the optima. Subsequently, the memory-intensive L-BFGS optimizer refines the final result. For this task, set 2 was used, and the epochs experimented were 10, 30, 50, 100, and 200 thousand. The results of varying the number of Adam epochs are shown in Fig. A.5. All reconstructed fields besides temperature were impacted, with $\varepsilon$ difference between the best and worst cases being about $4\times10^{-2}$, $1.8\times10^{-2}$, and $1.2\times10^{-2}$ for pressure, x-velocity, and y-velocity, respectively. The deviation in pressure follows an expected trend, exhibiting a negative correlation between $\varepsilon$ and epoch. Intriguingly, both velocity reconstructions demonstrated comparable $\varepsilon$ for small (10,000, 30,000) and large (200,000) epochs, but yielded larger $\varepsilon$ for intermediate (50,000, 100,000) epochs. This result suggests that 200,000 epochs is the optimal choice. However, $100\times10^3$ was chosen because it still fell within range of the U* standard deviation. An even smaller epoch could be chosen, but it was determined that there was no negative impact to a longer training duration. Furthermore, a longer epoch could assist PINN convergence for more complex problems (problems that typically converge slower).